\documentclass[twocolumn, amsmath,superscriptaddress, amsfonts,prb]{revtex4}
\usepackage{graphicx}
\begin{document}

\title{Semi-shunt field emission in electronic devices}
\author{V. G. Karpov}\email{victor.karpov@utoledo.edu}\affiliation{Department of Physics and Astronomy, University of Toledo, Toledo, OH 43606, USA}
\author{Diana Shvydka}\email{diana.shvydka@utoledo.edu}\affiliation{Department of Radiation Oncology, University of Toledo, Toledo, OH 43606, USA}

\date{\today}

\begin{abstract}
We introduce a concept of semi-shunts representing needle shaped metallic protrusions shorter than the distance between a device electrodes. Due to the lightening rod type of field enhancement, they induce strong electron emission. We consider the corresponding signature effects in photovoltaic applications; they are: low open circuit voltages and exponentially strong random device leakiness. Comparing the proposed theory with our data for CdTe based solar cells, we conclude that stress can stimulate semi-shunts' growth making them shunting failure precursors. In the meantime, controllable semi-shunts can play a positive role mitigating the back field effects in photovoltaics.
\end{abstract}
\maketitle

\begin{figure}[tb]
\includegraphics[width=0.40\textwidth]{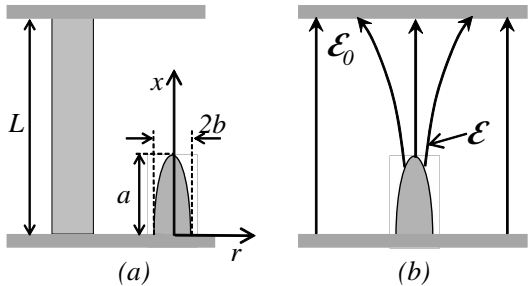}
\caption{(a) Sketch of a device with the full shunt (left) and semi-shunt (right). (b) Sketch of the electric field distribution in the semi-shunt proximity. The field ${\cal E}$ concentrates at its tip, similar to the lighting rod effect; the uniform field far away from the tip is relatively  weak, ${\cal E}_0\ll {\cal E}$.  \label{Fig:shunts}}
\end{figure}

The detrimental role of shunts is known in many technologies ranging from microelectronics to large area photovoltaics (PV). The most widely known, ohmic shunts are commonly attributed to metallic filaments forming conductive pathways between the device electrodes. In some cases, shunts exhibit the non-ohmic behavior described as weak micro-diodes. \cite{karpov2002,shvydka2003,karpov2004} Here we consider 'shunts to be' that are metallic in nature, but remain developed only partially, not connecting the device electrodes as illustrated in Fig. \ref{Fig:shunts}; we call them semi-shunts.

The existence of semi-shunts follows from the fact that full shunts develop by growth, and during that process their lengths are shorter than the inter-electrode distance, $a<L$. As an example, for a particular type of shunts related to metal whiskers, the existence of broad distribution of lengths is well established. \cite{fang2006,susan2013} A possible role of semi-shunts in electronics has never been explored. Here, we show that they can noticeably affect the device characteristics, and that their signature features can be used for diagnostic tests.

We model a semiconductor junction with the built in electric field ${\cal E}_0$ in a uniform layer between two electrodes.  It contains a semi-shunt modeled as a half of a prolate spheroid with semi-axes $a$ and $b\ll a$ shown in Fig. \ref{Fig:shunts}; assuming a cylinder shape leads to similar predictions. Fig. \ref{Fig:shunts} (b) illustrates a major physical effect by a semi-shunt: local electric field enhancement near the shunt tip similar to the well-known lightning-rod action.

\begin{figure}[bht]
\includegraphics[width=0.50\textwidth]{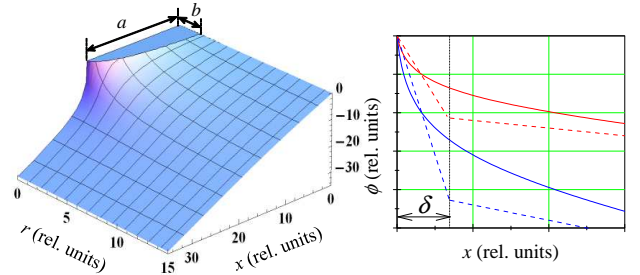}
\caption{Left: Electric potential  of Eq. (\ref{eq:poten}); the flat portion represents a quarter of metallic semi-shunt. Right: electric potential along $x$-axis for two different field strengths. Dashed lines show the piecewise  approximation of Eq. (\ref{eq:energy}).  \label{Fig:field}}
\end{figure}

A semi-spheroid on the electrode surface creates the same potential as the full prolate spheroid formed by the original and image half-spheroidal charges, \cite{landau1984,bat1964}
\begin{eqnarray}\label{eq:poten}
\phi &=& -\frac{{\cal E}_0x}{\Lambda}\left[\Lambda-\ln\left(\frac{\sqrt{1+\xi /a^2}+e}{\sqrt{1+\xi /a^2}-e}\right)+\frac{2e}{\sqrt{1+\xi /a^2 }}\right]\nonumber\\ \Lambda &\equiv & \ln\left(\frac{1+e}{1-e}\right)-2e,\quad
 e=\sqrt{1-\frac{b^2}{a^2}},
\end{eqnarray}
where $e$ is the eccentricity and $\xi$ is defined by
\begin{equation}\label{eq:xi}
x^2/(a^2+\xi )+r^2/(b^2+\xi )=1,
\end{equation}
$x$ and $r$ being respectively transversal (along the semi-shunt axis) and radial (parallel to the electrode) coordinates. The potential of Eq. (\ref{eq:poten}) is illustrated in Fig. \ref{Fig:field}.

The field concentration effect is described by the enhancement factor $\alpha$,
\begin{equation}\label{eq:field}
{\cal E}=\alpha {\cal E}_0\gg {\cal E}_0\quad \quad \alpha\equiv a^2/(b^2\Lambda )\gg 1.
\end{equation}
In what follows, we approximate the enhanced field region by the linear term in the expansion of Eq. (\ref{eq:poten}),
\begin{equation}\label{eq:energy}
\phi (r=0)=\left\{\begin{array}{lll}\quad 0\quad &{\rm when}& \quad 0<x<a,\\
-{\cal E}x \quad &{\rm when}&\quad a<x<a+\delta ,\\
-{\cal E}_0x \quad &{\rm when}&\quad a+\delta <x<L\end{array}\right.
\end{equation}
where
\begin{equation}\label{eq:alpha}
\delta\equiv a/\alpha\ll a.
\end{equation}
Note that ${\cal E}\delta ={\cal E}_0a$, i. e. the electric potential drop across the $\delta$-region equals that across the semi-shunt length in the host material. Note also that the image charges of the opposite electrode do not significantly affect the field in Eq. (\ref{eq:field}) as long as the semi-shunt tip remains at distance $L-a\gg \delta$ from it.

Corresponding to large $\alpha$ are small $\delta\lesssim 10$ nm (see Table \ref{tab:shuntpar} below) allowing efficient quantum tunneling through the enhanced field region. This understanding is consistent with the observation that tens of nanometers thin films form efficient tunnel junctions in PV. \cite{yamaguchi2005,siyu2011,hegedus1995} The exponentially enhanced electron emission rates  make semi-shunts effective field emission guns.\cite{sze,fursey2005}

\begin{figure}[tb]
\includegraphics[width=0.3\textwidth]{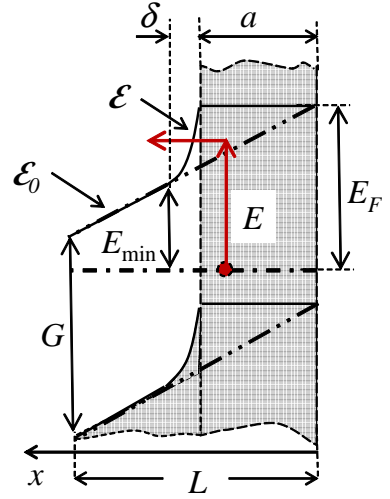}
\caption{The energy band diagram of a structure with semi-shunt. The solid horizontal domains show the conduction and valence band edges tangent to the semi-shunt. The solid lines at $x>a$ represent these edges at the semi-shunt axis (cf. Fig. \ref{Fig:field}). The double dot dash lines represent  them in the absence of semi-shunt. $E_F$ is the Fermi energy, $E$ is the activation energy, $G$ is the semiconductor forbidden gap. The arrows represent thermally activated tunneling. \label{Fig:band}}
\end{figure}

More specifically, consider the triangular barrier of width $\delta\lesssim 10$ nm and height $V_B=E_F-E_{\rm min}\lesssim 1$ eV in Fig. \ref{Fig:band}. Here  $E_{\rm min}=E_F-{\cal E}q\delta$ is the minimum excitation energy allowing the electron tunneling through the region of width $\delta$, $q$ is the elemental charge, and $E_F$ is the Fermi level. For any practical values of parameters, the characteristic tunneling exponent $S_{\rm min}\sim \sqrt{mV_B}\delta /\hbar\lesssim 10$ turns out to be much smaller than the exponent $V_B/kT$ of the probability of activation to the top of the barrier; hence, tunneling at $E_{\rm min}$ prevailing.

% allows the electric current density $j_T\sim qnv_F\exp(-S)$ where $n$ and $v_F$ are the electron concentration and Fermi velocity in the semi-shunt metal, and $q$ is the electron charge. For any practical values of parameters here, $j_T$  is much greater than the bulk current density $j_s= \sigma {\cal E}_0$ where $\sigma$ is the bulk conductivity.

Furthermore, tunneling is suppressed for electrons activated below $E_{\rm min}$ because their corresponding barriers are determined by the field ${\cal E}_0\ll {\cal E}$ with tunneling exponents $S\sim \alpha S_{\rm min}\gg S_{\rm min}$. Therefore, semi-shunts possessing large enough $\alpha$ will provide  saturation currents with activation energy close to $E_{\rm min}$.

%rather than its value of $E_F>E_{\rm min}$ in the absence of semi-shunts.

The above decrease in activation energy is 
\begin{equation}\label{eq:Vbi}E_F-E_{\rm min}={\cal E}_{0}aq=\min\{(V_{bi}-V)aq/L, \quad E_F\}\end{equation}
where $V_{bi}$ is the built-in potential, $V$ is the external bias; the Fermi energy limitation preserves the condition of activated saturation current. The corresponding current-voltage characteristic for not too negative $V$ [cf. Eq. (\ref{eq:Vbi})] can be presented in the form
\begin{equation}\label{eq:IVm}
I=I_0\left\{\exp\left[\frac{qa(V_{bi}-V)}{LkT}\right]\right\}\left[\exp\left(\frac{qV}{kT}\right)-1\right]-I_L
\end{equation}
where $I_L$ is the photocurrent, and
\begin{equation}\label{eq:I00}
I_0\equiv I_{00}\exp\left(-E_F/kT\right), \quad I_{00}=const,
\end{equation}
The criterion of large enough $\alpha$ will be given below.

\begin{figure}[tbh]
\includegraphics[width=0.36\textwidth]{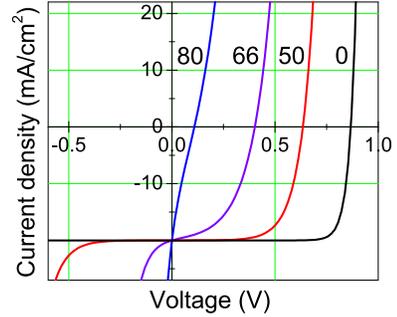}
\caption{Current voltage characteristics of Eq. (\ref{eq:IVm}) with parameters $V_{bi}=1.1$ eV, $q/kT=30$/V, $I_{00}=10^{-8}$ mA/cm$^2$, and $I_L=20$ mA/cm$^2$. Figures by the curves show percent values of the relative semi-shunt lengths $a/L$.  \label{Fig:IV}}
\end{figure}

Examples of such IV curves are shown in Fig. \ref{Fig:IV}. Their characteristic parameters are: the open circuit voltage,
\begin{equation}\label{eq:Voc}
V_{oc}\approx\frac{kTL}{q(L-a)}\ln\left(\frac{I_L}{I_0}\right)-\left(\frac{V_{bi}a}{L-a}\right),
\end{equation}
the open circuit and short circuit resistances,
\begin{equation}\label{eq:Roc}
R_{oc}=\frac{kTL}{qI_L(L-a)},\quad R_{sc}=R_{oc}\frac{I_L}{I_0}\exp\left(-\frac{qV_{bi}a}{kTL}\right).
\end{equation}
%and the short circuit resistance,
%\begin{equation}\label{eq:Rsc}
%R_{sc}=\frac{kTL}{qI_0(L-a)}\exp\left(-\frac{qV_{bi}a}{kTL}\right).
%\end{equation}

\begin{table}[t!]
\caption{Shunt independent parameters of a typical CdTe based solar cell.\cite{Sites,karpov2002} $D$ is the cell diameter, and $\rho$ is the sheet resistance; other parameters are defined in the text. The value of $V_{bi}$ greater than $V_{oc}$, but smaller than $G$ is assumed.}
  \begin{tabular}{| c | c | c | c | c | c | c | c | c | c |}
  \hline
    $L$, & $T$, & $m$, & $V_{oc}$, & $I_L$, &$V_{bi}$ & ${\cal E}_T$ & $\rho$,& $D$, \\
    $\mu$m & K & g & V & mA/cm$^2$  & V & kV/cm &Ohm /$\square$ &  cm \\ \hline
    $3$ & 300 & $10^{-27}$ & 0.8 & 20 & 1.1 & 260 & 10  & 1  \\ \hline
    %$1.5$ & 10 & &  &  &  &  &  \\  \hline
  \end{tabular}\label{tab:cell}
\end{table}

The following observations are worth mentioning.\\ (1) Semi-shunts create IV characteristics with decreased $V_{oc}$ yet not significantly affecting the short circuit current and flatness of IV curves in the proximity of $V=0$, unlike ohmic shunts. Such characteristics have been previously attributed to weak diodes, \cite{karpov2002,shvydka2003,karpov2004,roussillon2004}  the microscopic nature of which remained an open question. Our consideration here shows that the weak diode behavior can be caused by semi-shunts.\\(2) A rather insignificant effect of semi-shunts on $R_{oc}$ can be masked by other factors.\\ (3) $R_{sc}$ is exponentially affected by semi-shunts. The exponential dispersion in $R_{sc}$ reflecting semi-shunt randomness can be used for diagnostic purposes.\\ (4) Under negative bias, $V\lesssim V_{bi}-V_{oc}L/a$, IV curves exhibit exponential fall off. That feature can be suppressed by the limitation of Eq. (\ref{eq:Vbi}), i. e. $V>V_{bi}-E_FL/qa$ governed by other device parameters.

To establish the domain of applicability of the above results, we consider the case when the barrier width $\delta$ is not small enough to allow efficient tunneling at energy $E_{\rm min}$. In that case, the electrons overcome a barrier via thermally activated tunneling \cite{abakumov1991} at energy $E>E_{\rm min}$ in Fig. \ref{Fig:band}. The product of probabilities of activation and tunneling through the triangular barrier \cite{landau1977} determines the partial electric current,
\begin{equation}\label{eq:current}
I_E\propto\exp\left[-\frac{E}{kT}-\frac{4}{3}\frac{\sqrt{2m(E_F-E)}}{\hbar}\frac{E_F-E}{{\cal E}q}\right].
\end{equation}
It can be optimized to give the most efficient activation energy $E$ and its corresponding barrier width,
\begin{equation}\label{eq:opten}
E=E_F-\left(\frac{\hbar {\cal E}q}{\sqrt{8m}kT}\right)^2,\quad x_T=\frac{\hbar ^2{\cal E}q}{8m(kT)^2}.
\end{equation}
This yields
\begin{equation}\label{eq:current1}
\frac{I_T}{I_0}=\exp\left(\frac{{\cal E}}{{\cal E}_T}\right)^2=\exp\left(\frac{{\cal E}_0}{{\cal E}_{0T}}\right)^2=\exp\left(\frac{V_{bi}-V}{V_{0T}}\right)^2\end{equation}
with
\begin{equation}\label{eq:ET}
{\cal E}_T\equiv\frac{\sqrt{24m(kT)^3}}{\hbar q}, \quad {\cal E}_{0T}=\frac{{\cal E}_{T}}{\alpha}, \quad V_{0T}={\cal E}_{0T}L.
\end{equation}

The condition $x_T=\delta$ determines the maximum field and current,
\begin{equation}\label{eq:Ec}
{\cal E}_{0\delta}\equiv \frac{8m(kT)^2 a}{\hbar ^2q\alpha ^2}, \quad \left(I_{T}\right)_{max}=I_0\exp\left[\left(\frac{{\cal E}_{0\delta}}{{\cal E}_{0T}}\right)^2\right].
\end{equation}
Finally, the inequality $(I_T)_{max}\gg I_0$ translates into the limitation on the enhancement factor,
\begin{equation}\label{eq:alphamax}
\alpha \ll \alpha _c\equiv \sqrt{\frac{8kTma^2}{3\hbar ^2}}.
\end{equation}
Our results in Eqs. (\ref{eq:Vbi})-(\ref{eq:Roc}) belong to the region of $\alpha \gg \alpha _c$. Using thr parameters from Table \ref{tab:cell} yields $\alpha _c \approx 1000$.

In the region of $\alpha \ll\alpha _c$, Eq. (\ref{eq:current1}) leads to the IV characteristic
\begin{equation}\label{eq:IVmm}
I=I_0\left\{\exp\left[\frac{(V_{bi}-V)^2}{V_{0T}^2}\right]\right\}\left[\exp\left(\frac{qV}{kT}\right)-1\right]-I_L.
\end{equation}
Based on the latter, it is straightforward to derive closed form equations for $V_{oc}$, $R_{oc}$, and $R_{sc}$, analogous to that in Eqs. (\ref{eq:Voc})-(\ref{eq:Roc}); we skip them here as they present much weaker effects than that for the region of $\alpha\gg \alpha _c$.

Table \ref{tab:shuntpar}, illustrates various parameters related to four different semi-shunts in a cell described in Table \ref{tab:cell}. The cases \#1 and \#4 belong to the region of weak semi-shunts, $\alpha \ll \alpha _c$, while the cases \#2 and \#3 present significant semi-shunts with $\alpha\gg \alpha _c$. Its is worth noting that the enhancement factor (i. e. the aspect ratio) rather than the shunt length is the parameter that determines the significance of semi-shunts.
\begin{table}[bht]
\caption{Examples of the characteristic fields, voltages, and other parameters for four different semi-shunts.}
  \begin{tabular}{|c| c | c | c | c | c | c | c | c | c | c |}
  \hline

    \# &$a$, & $b$, & $\alpha$ & ${\cal E}_{0T}$, & $\delta$, &${\cal E}_{0\delta}$ & $V_{0T}$,& $V_{oc}$,  \\
     & $\mu$m & nm &  & kV/cm & nm  & kV/cm & V &  V \\ \hline
    1 &0.5 & 10 & 496 & 0.52 & 1 & 1.6 & 0.48  & 0.83 \\ \hline
    2 & 1.0 &10 &2347 & 0.11 & 0.42 & 0.15 &0.042 &0.65  \\ \hline
    3 &$1.5$ & 10 & 4787& 0.054 & 0.3 & 0.035 & 0.001 & 0.57  \\  \hline
    4 & $1.5$ & 100 &132 & 1.97 & 11.3 & 69 & 0.6 &0.85 \\ \hline
  \end{tabular}\label{tab:shuntpar}
\end{table}

We have verified some of the above predictions using our database of measurements of CdTe based solar cells described in Ref. \onlinecite{karpov2002}. In interpreting the data, it was taken into account that every cell can have a variety of different semi-shunts contributing to the cell's $R_{sc}$. Because they are connected in parallel, the cell reciprocal $R_{sc}$ can be considered as a sum of large number of random terms representing individual semi-shunt's reciprocal $R_{sc}^{-1}$'s. According to the cental limit theorem, \cite{dudley1999} such sums are Gaussian random quantities.

\begin{figure}[tb]
\includegraphics[width=0.50\textwidth]{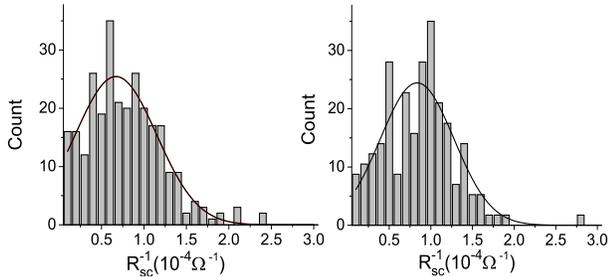}
\caption{Statistics of the short circuit conductances $R_{sc}^{-1}$ in the range from 26 Ohm to 15 kOhm, extracted from a database of 1020 CdTe based solar cells described in Ref. \onlinecite{karpov2002}, as grown (left) and after 28 days under light soak (right). The solid lines represent Gaussian fits. \label{Fig:Rsc}}
\end{figure}

The Gaussian distributions of $R_{sc}^{-1}$ were found indeed for the database of fresh and light soaked cells as illustrated in Fig. \ref{Fig:Rsc}. The observed correlation between $V_{oc}$ and $R_{sc}$ was not strong, which we attribute to the role of other factors affecting $V_{oc}$.

In addition, we found that the stressed cells exhibit on average $\approx 30$\% higher $R_{sc}^{-1}$ than the fresh cells. Simultaneously, the distribution width shrinks by $\sim 20$\% mostly due to dwindling the region of non-leaky cells. As related to semi-shunt lengths through Eq. (\ref{eq:Roc}), the observed decrease in $R_{sc}$ translates in the growth rate $da/dt\sim 0.001$ \AA/s, which is not unreasonable, being by two orders of magnitude below that of the extensively growing metal whiskers.\cite{fang2006,susan2013}

Finally, we have occasionally observed IV curves exhibiting reversible fall off at negative voltages ranging from $V\sim -2$ V to $V\sim -12$ V.  Stronger biasing and multiple cycling along the reversible fall off portion of IV often led to irreversible damage.  This is in general agreement with the above developed semi-shunt theory. For example the fall off at V=-3 V corresponds to the semi-shunt length $a\approx 0.2L$.

Since the semi-shunt growth kinetics fall beyond the current scope, here, we limit ourselves to pointing again to the analogy with the lightning rod effect. It suggests the possibility of local electric discharges detrimental to device stability; hence, semi-shunts serving as precursors of local device failures.

A comment is in order regarding the case of strong field emission by semi-shunts. Depending on the host parameters, it could create space charge limited currents (SLC). The related nontrivial questions about the multi-dimensional SLC in semiconductors (not solved enough even for vacuum plasma \cite{luginsland2002}), SLC under the condition of nano - and micro-structures, etc., remain beyond this paper frameworks.

The above consideration has revolved around detrimental effects of semi-shunts, such as $V_{oc}$ degradation and device leakiness. On the other hand, semi-shunts can be benign when they develop in and remain limited to the region of back contact. Such semi-shunts will provide the tunneling mechanism of overcoming the known detrimental back barrier (back field) effect. \cite{Sites, roussillon2004} This opens a venue of purposely engineering the benign semi-shunts at device back surfaces, through, e. g., use of suitable metals and back contact recipes.

In conclusion, we have introduced the concept of semi-shunts as partial metal protrusions and described their effects on device performance. Their signature features, such as the weak diode type of IV characteristics and exponential dispersion of short circuit resistances, can be used for device screening against complete shunting failure. In addition, purposely engineering semi-shunts on device back contact can help improve device performance by overcoming the back field effects.

This work was performed under the auspice of the NSF award No.  1066749.

%(1+\xi/a^2)^0.5=(0.00125*((i^2+j^2+396)+sqrt((i^2+j^2-404)^2+4*(400*j^2+4*i^2-40))))^(1/2)
%
%-i*(4-(ln(((0.00125*((i^2+j^2+396)+sqrt((i^2+j^2-404)^2+4*(400*j^2+4*i^2-40))))^(1/2)+0.995)/((0.00125*((i^2+j^2+396)+sqrt((i^2+j^2-404)^2+4*(400*j^2+4*i^2-40))))^(1/2)-0.995))-1.99/(0.00125*((i^2+j^2+396)+sqrt((i^2+j^2-404)^2+4*(400*j^2+4*i^2-40))))^(1/2)))
%
%-i*(4-(ln(((0.00125*((i^2+j^2+396)-sqrt((i^2+j^2-404)^2+4*(400*j^2+4*i^2-40))))^(1/2)+0.995)/((0.00125*((i^2+j^2+396)-sqrt((i^2+j^2-404)^2+4*(400*j^2+4*i^2-40))))^(1/2)-0.995))-1.99/(0.00125*((i^2+j^2+396)-sqrt((i^2+j^2-404)^2+4*(400*j^2+4*i^2-40))))^(1/2)))
%
%-x*(1-(ln((x+0.99)/(x-0.99))-1.98/x)/(ln(199)-1.98))


\begin{thebibliography}{99}
\bibitem{karpov2002}V. G. Karpov, A. D. Compaan, and D. Shvydka, Appl. Phys. Lett.  {\bf 80}, 2002, pp. 4256-4258.
\bibitem{shvydka2003}D. Shvydka, V. G. Karpov and A. D. Compaan, Appl. Phys. Lett. {\bf 82}, 2157 (2003).
\bibitem{karpov2004} V. G. Karpov, A. D. Compaan, and D. Shvydka, Phys. Rev B {\bf 69}, 045325 (2004).
\bibitem{fang2006}T. Fang, M. Osterman, and M. Pecht, Microelectron. Reliab. {\bf 46}, 846 (2006).
\bibitem{susan2013} D. Susan, J. Michael, R. P. Grant, B. McKenzie, and W. G.
Yelton, Metall. Mater. Trans. A {\bf 44}, 1485 (2013).
\bibitem{landau1984} L. D. Landau and E. M. Lifshitz, \emph{Electrodynamics of Continuous Media} (Pergamon, Oxford, New York, 1986).
\bibitem{bat1964}V. V. Batygin and I. N. Toptygin, Problems in Electrodynamics, Academic, London, (1964) p. 49.

\bibitem{yamaguchi2005}M. Yamaguchi, T. Takamoto, K. Araki, N. Ekins-Daukes, Solar Energy, {\bf 79},  78 (2005).
\bibitem{siyu2011} Lu Siyu and Qu Xiaosheng, Journal of Semiconductors, {\bf 32}, 112003 (2011).
\bibitem{hegedus1995}S. S. Hegedus, F. Kampas, and J. Xi, Appl. Phys. Lett. {\bf 67}, 813 (1995).
\bibitem{sze}S. M. Sze, {\it Physics of Semiconductor Devices}, Wiley, New York 1981.
\bibitem{fursey2005}G. Fursey, {\it Field emission in vacuum microelectronics}, Kluwer Academic/Plenum Publishers, New York 2005.
\bibitem{Sites}B. E. McCandless and J. R. Sites, in {\it Handbook of Photovoltaic Science and Engineering}, p. 617, Edited by A. Lique and S.
Hegedus, Wiley 2003.
\bibitem{dudley1999}R. M. Dudley, {\it Uniform Central Limit Theorems}, Cambridge, Academic (1999).
\bibitem{roussillon2004}Y. Roussillon, V. G. Karpov, D. Shvydka, J. Drayton, and A. D. Compaan, J. Appl.  Phys., {\bf 96}, 7283 (2004).
\bibitem{abakumov1991}V. N. Abakumov, V. I. Perel, I. N. Yassievich, {\it Nonradiative Recombination in Semiconductors} (North Holland, 1991).
\bibitem{landau1977} L. D. Landau and E. M. Lifshitz, {\it Quantum Mechanics. Nonrelativistic theory}. Pergamon Press, New York 1977.
\bibitem{luginsland2002}J. W. Luginsland, Y. Y. Lau, J. Umstattd, and J. J. Watrous, Physics of Plasmas, {\bf 9}, 2371 (2002). 




\end{thebibliography}
\end{document}